\pdfoutput=1

\documentclass[aps,prl,twocolumn,superscriptaddress]{revtex4}
\usepackage{graphics}
\usepackage{epstopdf}
\usepackage{graphicx}% Include figure files
\usepackage{dcolumn}% Align table columns on decimal point
\usepackage{bm}% bold math
\usepackage{amsmath}

\newcommand{\comm}[1]{{}}

\newcommand{\be}{\begin{equation}}
\newcommand{\ee}{\end{equation}}
\newcommand{\beq}{\begin{eqnarray}}
\newcommand{\eeq}{\end{eqnarray}}

\def\nue{\mathrel{{\nu_e}}}
\def\numu{\mathrel{{\nu_\mu}}}
\def\nutau{\mathrel{{\nu_\tau}}}
\def\nux{\mathrel{{\nu_x}}}

\def\barnue{\mathrel{{\bar \nu}_e}}
\def\barnumu{\mathrel{{\bar \nu}_\mu}}
\def\barnutau{\mathrel{{\bar \nu}_\tau}}

\def \lta {\mathrel{\vcenter{\hbox{$<$}\nointerlineskip\hbox{$\sim$}}}}
\def \gta {\mathrel{\vcenter{\hbox{$>$}\nointerlineskip\hbox{$\sim$}}}}

\def\t13{\mathrel{{\theta_{13}}}}
\def\y12{\mathrel{{\tan^2 \theta_{12}}}}
\def\c2{\mathrel{{\chi^2 }}}

\newcommand{\n}{neutrino}
\newcommand{\ns}{neutrinos}

\newcommand{\sn}{supernova}
\newcommand{\sne}{supernovae}

\newcommand{\bh}{black hole-forming}
\newcommand{\nts}{neutron star-forming}

\newcommand{\ck}{Cherenkov}

\newcommand{\sk}{SuperKamiokande}

\newcommand{\nsf}{NSFCs}
\newcommand{\bhf}{DBHFCs}

\begin{document}

%\preprint{}

\title{Revealing local failed supernovae with neutrino telescopes}

\author{Lili Yang}
\affiliation{Arizona State University, Tempe, AZ 85287-1504}%

\author{Cecilia Lunardini}
 %\email{Cecilia.Lunardini@asu.edu}   
\affiliation{Arizona State University, Tempe, AZ 85287-1504}%
\affiliation{RIKEN BNL Research Center, Brookhaven National Laboratory, Upton, NY 11973}

%\date{\today}% It is always \today, today,
             %  but any date may be explicitly specified
 
\begin{abstract}
We study the detectability of \n\ bursts from nearby direct \bh\ collapses (failed \sne) at Megaton detectors. Due to their  high energetics, these bursts could be identified -- by the time coincidence of $N\geq 2$ or $N\geq 3$ events within a $\sim 1$ s time window -- from as far as $\sim 4-5$ Mpc away. 
This distance encloses several  \sn-rich galaxies, so that failed \sn\ bursts could be detected at a rate of up to one per decade, comparable to the expected rate of the more common, but less energetic, \nts\ collapses.  Thus,  the detection of a failed \sn\ within the lifetime of a Mt detector is  realistic. It might give the first evidence of direct black hole formation, with important implications on the physics of this phenomenon.
\end{abstract}                            
 
% PACS, the Physics and Astronomy
                             % Classification Scheme.
%\keywords{Suggested keywords}%Use showkeys class option if keyword
                              %display desired
\maketitle
  
%%%%%%%%%%%%%%%%%%%%%%%%%%%%%%%  
%\section{Introduction}

The gravitational collapse of a stellar core is one of the most extreme phenomena in our universe. There, matter is pushed to its limits of density, and most of the energy is emitted by a non-electromagnetic form of radiation, the \ns, rather than in the final explosion (supernova) that often follows the collapse.  

Neutrinos are true tracers of core collapse.  Due to their long mean free path, they give a direct image of the outskirts of the collapsed core.  Furthermore, they are the only emission -- together with gravitational waves --  that always accompanies a collapse!  Indeed,
it is predicted
  that  10-20\% of  collapses  directly generate a black hole \cite{Woosley:2002zz}, with a brief and strong phase of \n\ emission, and no explosion \cite{Liebendoerfer:2002xn,Sumiyoshi:2006id,O'Connor:2010tk}.  For these \emph{ failed supernovae}, the star simply disappears from the sky, leaving the \n\ burst as a unique messenger of the event. 

At present, the  detection of \sn\ \n\ bursts  is 
still limited by long waiting times,  
as current detectors -- of ${\mathcal O}(10)$ kt mass --  can only capture the 1-3 bursts per century in our galactic neighborhood \cite{Arnaud:2003zr,Ando:2005ka,Kistler:2008us}. Upcoming Mt scale detectors will start to overcome the time barrier: for the common, \nts\ collapses (which have an accompanying explosion), they have a volume of sensitivity of 1-2 Mpc radius \cite{Ando:2005ka}, where about  $\sim 1$ collapse per decade is predicted  \cite{Ando:2005ka}.
By applying a 10-20\% fraction, this translates into about 1-2 detections of  failed \sn\ bursts per century, still discouraging for an experiment lifetime of a few decades.

In fact, however, certain factors enhance the detectability of a \n\ burst from failed \sne.
First, the higher \n\ luminosity and average energy of failed \sne\ corresponds to a larger distance of sensitivity, a distance that    happens -- as will be seen here -- to be just enough to bring within the range of observability several major, \sn-rich galaxies located  3-4 Mpc away.  This fortunate circumstance can boost the expected detection rate significantly, similarly to what was discussed  for the diffuse \sn\ \n\ flux \cite{Lunardini:2009ya}.
 Furthermore, the shorter duration of a failed \sn\ burst (0.5-1 s) makes it easier to identify: the time coincidence of two \n\ events within $\sim 1$ s or so might be sufficient for discrimination against background.

 The fact that detecting individual \n\ bursts from failed \sne\ is realistic, with a Mt detector, implies the potential to reveal -- possibly for the first time -- the direct collapse of a star into a black hole, with several implications on the physics of this transition, such as the rate of accretion of matter on the collapsed core, the equation of state of nuclear matter, etc..  
 Here we elaborate on the idea of the enhanced detection rate of failed \sn\ bursts, and discuss its implications. 

%%%%%%%%%%%%%%%%%%%%%%%%%%%%%%%  
%\section{Failed supernovae:  neutrino signatures and nearby rate}

Failed \sne\ (or direct black hole forming collapses, \bhf) are predicted to originate from stars  with mass above $M_{min}\sim 25-40~M_{\odot}$ (with $M_{\odot}$  the mass of the Sun) \cite{Woosley:2002zz,O'Connor:2010tk}, corresponding to 9-22\% of all collapsing stars (see, e.g., \cite{Lunardini:2009ya}). 
Numerical simulations \cite{Liebendoerfer:2002xn,Sumiyoshi:2006id,Sumiyoshi:2007pp,Fischer:2008rh,Sumiyoshi:2008zw,Nakazato:2008vj} indicate that their \n\ burst lasts  $\sim 1$ s or less, and has up to $L\sim 5 \cdot 10^{53}$ ergs luminosity, due to the rapid
contraction of the newly formed protoneutron star preceding the black hole formation.
   The produced electron \ns\ and antineutrinos, $\nue$ and $\barnue$, have especially high luminosity, $L_{0 e} \simeq L_{0 \bar e}\sim 10^{53}$ ergs, due to the high rate of electron and positron captures on nuclei. Their  average energy can reach $E_{0 e}\simeq E_{0 \bar e}\sim 20-24$ MeV.   
    
   Due to oscillations in the star,    the $\barnue$ flux in a detector is  an admixture of the unoscillated flavor fluxes: $F_{\bar e}=\bar  p F^0_{\bar e} + (1-\bar p) F^0_{x}$, where $x$ indicates the non-electron species, $\nux=\numu,\barnumu,\nutau,\barnutau$, and  $\bar p$ is the $\barnue$ survival probability \cite{Dighe:1999bi}. Following \cite{Nakazato:2008vj,Lunardini:2009ya,Keehn:2010pn}, we consider $\bar p=0 - 0.68$ and give results for the energy-independent, limiting case of $\bar p=0.68$, unless otherwise specified. 
  We take the \n\ fluxes from  fig. 5 of \cite{Nakazato:2008vj} \footnote{We used a fourth order polynomial interpolation of the logarithmic spectra in \cite{Nakazato:2008vj}. 
  %The associated error  is difficult to estimate, however 
  We have verified that our positron spectra look natural, depend only weakly on the interpolation order, and are close to those in \cite{Nakazato:2008vj}.  }, for the Shen et al. equation of state of nuclear matter.   This set of flux and oscillation parameters maximizes $F_{\bar e}$ \cite{Nakazato:2008vj}, and so it is adequate to estimate the maximum potential of detection of failed \sne.  
   
 For comparison,  we also model bursts from \nts\ collapses (\nsf). These last 10-20 s and have typical parameters $L\sim 3 \cdot 10^{53}$ ergs, $L_{0 \bar e}\sim L_{0 x}\sim 0.5 \cdot 10^{53}$ ergs, $E_{0 \bar e} \sim 15$ MeV, $E_{0 x} \sim 18$ MeV.  The spectra of the produced \ns\ in each flavor typically have the form of a power-law times an exponential \cite{Keil:2002in}. We restrict to the case in which $\bar p$ has the same value for the two collapse types  \cite{Nakazato:2008vj,Keehn:2010pn}. 

\begin{figure}[htbp]
%\begin{figure}[t]
  \centering
 \includegraphics[width=0.45\textwidth]{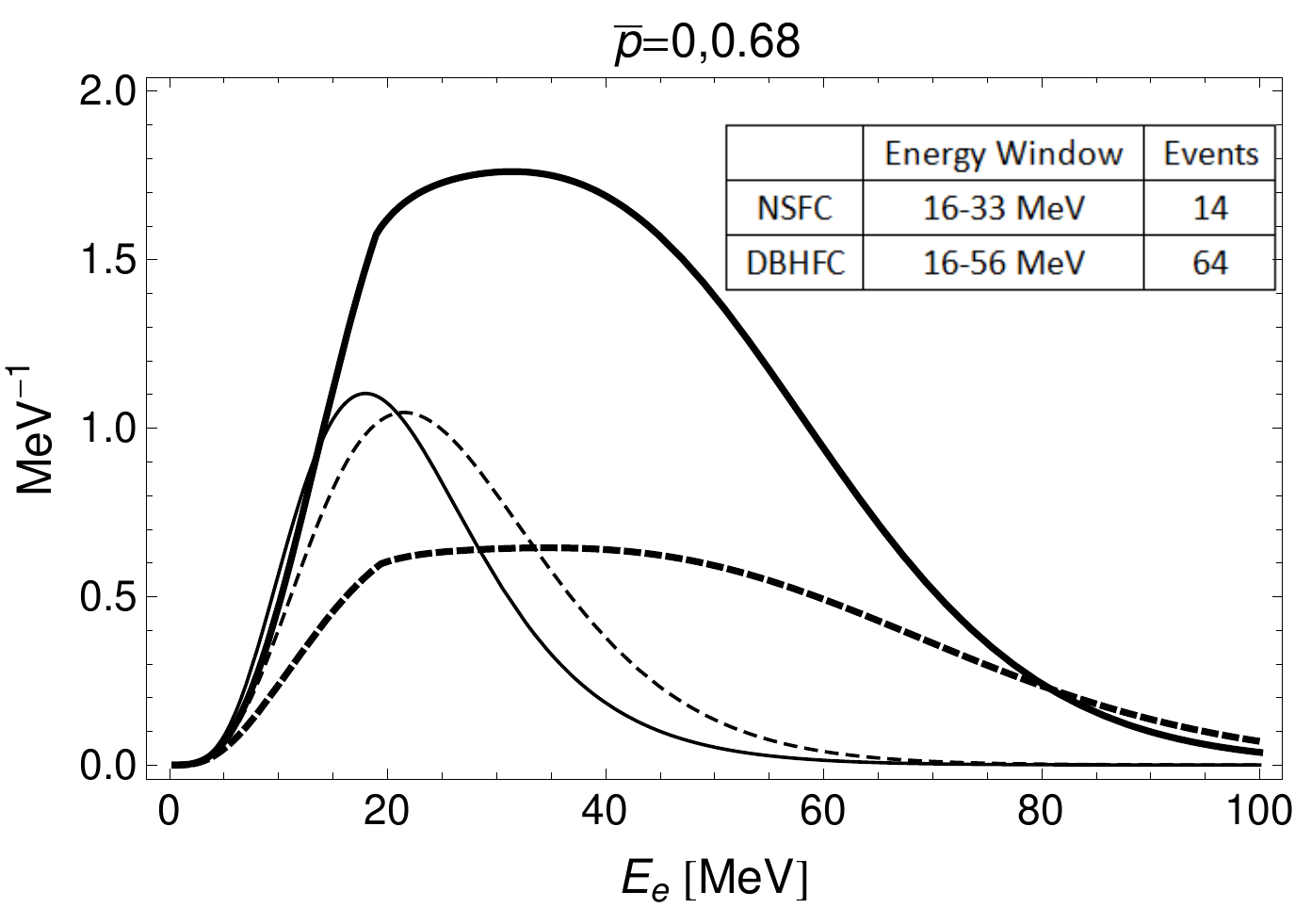}
   \caption{Positron energy spectra at a 1 Mt water \ck\ detector from a \nts\ forming collapse (thin curves) and a \bh\ forming collapse (failed supernova, thick curves), at distance $D=1$ Mpc.   Dashed curves: $\bar p=0$; solid: $\bar p=0.68$.  The Shen et al. equation of state is used  for the failed \sn\  \cite{Nakazato:2008vj}. Integrated numbers of events are also given for $\bar p=0.68$ and realistic energy windows of detection (see text).  }
\label{spectraevents}
\end{figure}

Let us consider the response of a 1 Mt water \ck\ detector \cite{Jung:1999jq,Nakamura:2003hk,deBellefon:2006vq}  to a \n\ burst. The dominant detection reaction is inverse beta decay, $\barnue + p \rightarrow n + e^+ $, which we model as in \cite{Strumia:2003zx}.    The expected positron spectra for the two collapse types are shown in fig. \ref{spectraevents}.  The higher energetics of a failed \sn\ is evident in the figure. 

An experiment looks for inverse beta decay events  within fixed time and energy windows designed to maximize the signal to background ratio \cite{Ikeda:2007sa}.   Typical time windows could be $\Delta t=10$ s and $\Delta t=1$ s for \nts\ collapses and failed \sne\ \footnote{We neglect time delay effects due to the \n\ mass \cite{Piran:1981zz}. 
  This is adequate for masses $m_\nu \lta 0.7$  eV.  }. The energy windows 
are limited by background at low energy; a threshold of about 16 MeV in positron energy seems realistic  \cite{Ikeda:2007sa}.  The windows could be defined as including at least $80\%$ of the events above this threshold: we find the intervals $E_e = 16 - 33$ MeV and $E_e = 16 - 56$ MeV for \nsf\ and \bhf\ respectively.  A \n\ burst is identified (``detected") if $N \geq N_{min}\sim 2-3$ events are observed in the energy window with time separation less than $\Delta t$.  
Note that  the number of events due to a failed \sn\ increases with increasing $\bar p$ (fig. \ref{spectraevents}), i.e., with larger survival of the more luminous original $\barnue$ component.  It can be as large as $\mu(D)\simeq 64 ({\rm 1~ Mpc}/D)^2$, up  to five times larger than that from a NSFC.  Therefore, 2 (3) events are expected from a failed \sn\ as far as $D \sim 6$ Mpc ($D \sim 5$ Mpc).

\begin{figure}[htbp]
%\begin{figure}[t]
  \centering
 \includegraphics[width=0.4\textwidth]{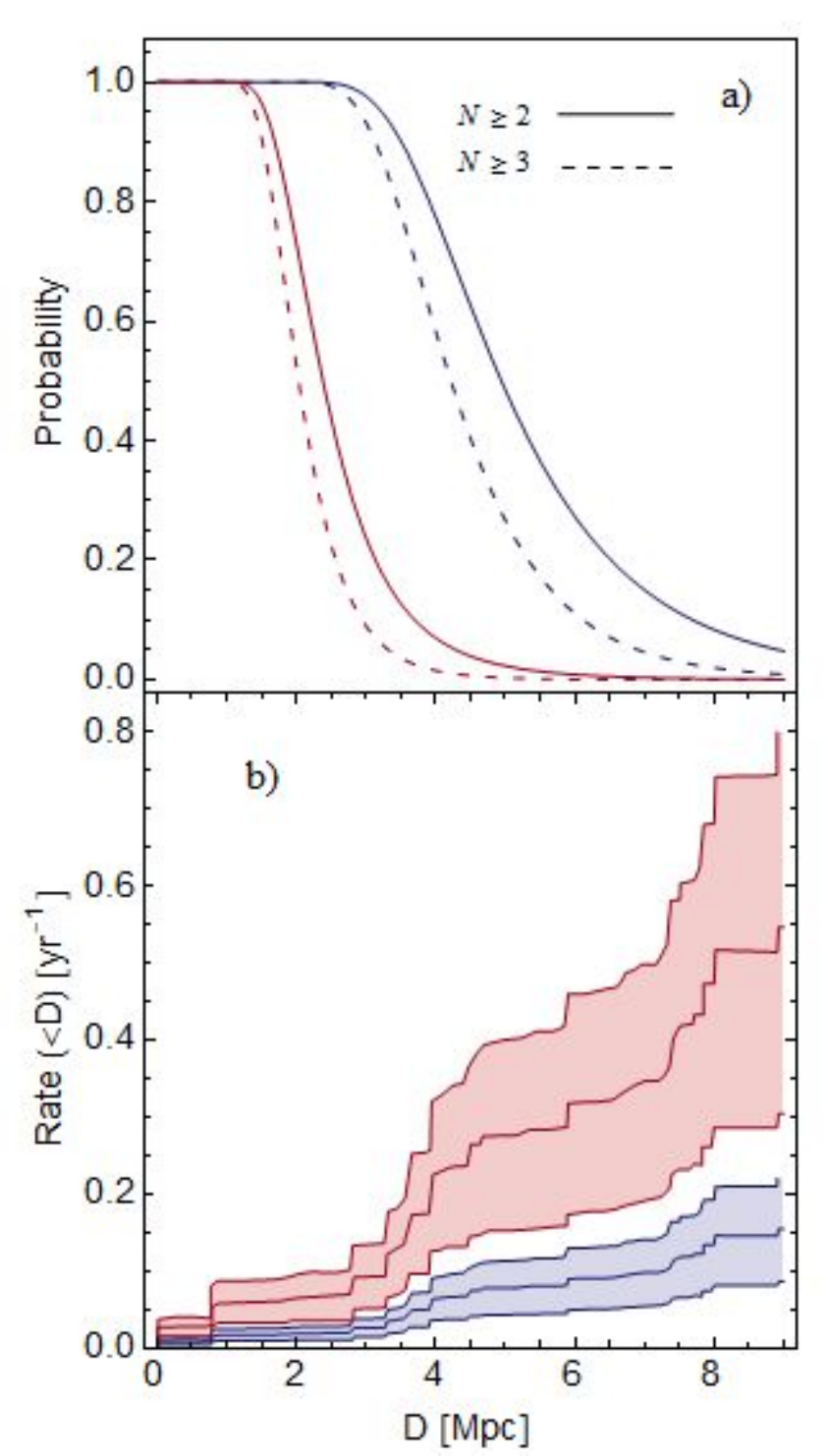}
   \caption{a): The Poissonian probabilities to detect $N\geq 2$ and $N \geq 3$  events at a 1 Mt water \ck\ detector for \nts\ collapses (lower curves, red) and \bh\  collapses (upper curves, blue).  The results of fig. \ref{spectraevents} with $\bar p=0.68$ are used.  b): the rates of the two collapse types (the lower curves refer to failed \sne)  within a radius $D$ from Earth,   with their uncertainties. They are taken from \cite{Ando:2005ka} with a fraction $f_{BH}=0.22$ of  failed \sne.   }
\label{ccrates}
\end{figure}

Given the ``true" number of  events, $\mu$, the probability of detection of a burst (i.e., $N \geq N_{min}$ positrons observed) in the detector is given by the Poisson distribution: 
\begin{equation}
P(N_{min},D)=\sum_{n=N_{min}}^{\infty} \frac{\mu^{n}(D)}{n!} e^{-\mu(D)}~
\label{prob}
\end{equation}
\cite{Kistler:2008us}. It is shown in fig. \ref{ccrates}a) as a function of the distance $D$ for $N_{min}=2,3$.  The figure confirms the expectation of a larger range of sensitivity to failed \sne, with a probability of detection as large as 0.8 for $D=D_{s}\simeq 4-4.5$ Mpc, which can be thus considered a typical distance of sensitivity. 
The corresponding distance for a \nts\ collapse is $D_{s}\simeq 2-2.5$ Mpc. 

We now come to the key point of this work: how the increased distance of sensitivity allows to probe a region of high core collapse rate.  
Fig. \ref{ccrates}b) gives the nearby rates of the two types of collapses, $R_{BH}$ and $R_{NS}$, within a distance $D$, with their uncertainty.  They are derived from the collapse rate in \cite{Ando:2005ka} (which is obtained from  a catalog of galaxies \cite{Karachentsev:2004dx} with conversion factors between luminosities and core collapse rates \cite{Cappellaro:1999qy}), under the assumption of a constant, distance-independent, ratio $f_{BH}=0.22$ of failed \sne\ \footnote{This assumption is necessarily tentative, as no data exist about the distribution of failed \sne.}. These rates  are higher than the cosmological average \cite{Ando:2005ka}, and actual supernova observations favor 
 an even higher rate \cite{Kistler:2008us}. Therefore, our results based on fig. \ref{ccrates}b) are conservative. 

Fig. \ref{ccrates}b) clearly shows the rapid increase of the rates between 3 and 4 Mpc, due to the presence of several galaxies (mainly IC 342, NGC 2403, M 81, M 82, NGC 4945 \cite{Ando:2005ka}),  in this interval of distance. This is well within the range of sensitivity for failed \sne, but only marginally accessible for the less luminous \nsf. 
Within the typical distance of sensitivity, $D_s$, fig. \ref{ccrates}b) gives a rate of  $\sim 0.04 - 0.10~{\rm yr^{-1}}$ for failed \sne, and $\sim 0.07 - 0.14~{\rm yr^{-1}}$ for \nts\ collapses. The two rates are comparable, showing that the increased distance of sensitivity for failed \sne\ compensates in part for their rarity.

%%%%%%%%%%%%%%%%%%%%%%%%%%%%%%%  
%\section{Results: expected rate of bursts in a detector}

One can calculate the expected rate of detections of bursts from \bhf\ within a distance $D$ \cite{Kistler:2008us}:
\begin{equation}
R^{det}_{BH}( N_{min}, D) = \sum_{i, ~D_i \leq D} \Delta R_{BH,i} P(N_{min}, D_i)~.
\label{exprates}
\end{equation}
  The sum is over bins of distance, with $D_i \leq D$, and $ \Delta R_{BH,i}$ is the failed \sn\ rate in each bin, so that $\sum_{i, ~D_i \leq D} \Delta R_{BH,i}  = R_{BH}(D)$.   An expression analogous to eq. (\ref{exprates}) holds for the rate of detections of  \nsf, $R^{det}_{NS}$. 

  \begin{figure}[htbp]
%\begin{figure}[t]
  \centering
 \includegraphics[width=0.45\textwidth]{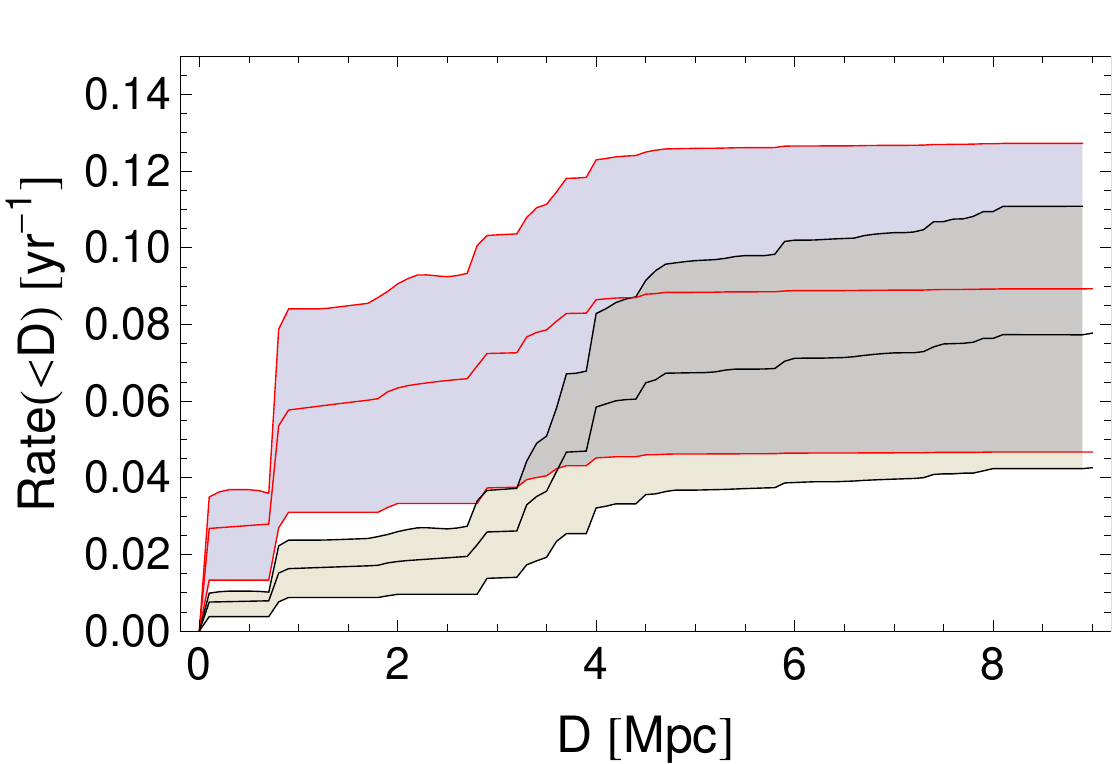}
   \caption{The expected rates of detections of \n\ bursts that originate within a radius $D$ from Earth, as functions of $D$, for \nts\  and \bh\  collapses (upper and lower shaded regions).   All parameters are as in fig. \ref{ccrates}  for the case  $N\geq 2$.   }
\label{burstrates}
\end{figure}
Fig. \ref{burstrates} gives $R^{det}_{BH}$ and $R^{det}_{NS}$ as functions of the distance. 
Naturally, for each \sn\ type the detection rate follows the collapse rate for $D \ll D_{s}$; it then flattens for larger distances, reflecting the suppression due to  the small detection probability (fig. \ref{ccrates}a).  This flattening occurs around 4 Mpc for \nsf, 
and at $\sim 8-9$ Mpc for \bhf.  Depending on the normalization of the collapse rate, the detection rates for the two collapse types reach $R^{det}_{NS} \sim 0.05 - 0.13~{\rm yr^{-1}}$  and $R^{det}_{BH}\sim 0.04 - 0.11~{\rm yr^{-1}}$. 

Thus, failed \sne\ have a chance to be detected within the lifetime of an experiment.  Due to their contribution, the total rate of burst detections could be twice as large as previously estimated, with a maximum of about 2 detections per decade.

%%%%%%%%%%%%%%%%%%%%%%%%%%%%%%%  
%\section{Background considerations}

Typically, expected detection rates are considered promising if they exceed the corresponding background rates, so that an observed burst can be attributed to a \sn\ with substantial likelihood. 
Assuming that correlated events can be identified and subtracted \cite{Ikeda:2007sa}, the background is given by accidental coincidences of uncorrelated events within the energy and time windows.  
By rescaling the \sk\ measurements \cite{Malek:2002ns,iidathesis} to a Mt mass, we find the rates of uncorrelated events to be  $\lambda= 1855~{\rm yr^{-1}}$ ($\lambda= 680~{\rm yr^{-1}} $ ) in the energy window for \bhf\ (\nsf).

The rate of coincidence of two (three) such uncorrelated events in the time window is (for $\lambda \Delta t \ll 1$)  $\omega_{2}\simeq \lambda^2  \Delta t  $ ($\omega_{3}\simeq \lambda^3 \Delta t^2 $) \cite{coxbook}. 
For failed \sne\ ($\Delta t = 1$ s) we find $\omega_{2} \simeq 0.10~{\rm yr^{-1}}$ and  $\omega_{3}\simeq 6.4 \times 10^{-6}~{\rm yr^{-1}}$.  
The same quantities for the \nsf\ time window are $ \omega_{2}\simeq   0.15~{\rm yr^{-1}}$   and  $\omega_{3}\simeq 3.1 \times 10^{-5} ~{\rm yr^{-1}}$.  For both collapse types, the background doublet rate is comparable to or only slightly higher than the burst rate, so two observed positron events might be sufficient to claim a \sn\ detection, depending on the details of the experimental setup, and three events should give practically certain identification.  For a \nts\ collapse, the identification will probably be confirmed by the observation of the \sn\ explosion at telescopes, unless  obscuration is substantial.  For failed \sne, one would have to rely entirely on \ns, or, possibly, on the  concident detection of gravitational waves, or on establishing the disappearance of the star \cite{Kochanek:2008mp}.
%detection in coincidence of \ns\ and gravitational waves.  Therefore, $N_{min}=3$ might be required for a robust detection.  

%%%%%%%%%%%%%%%%%%%%%%%%%%%%%%%  
%\section{Discussion and summary}

  \begin{figure}[htbp]
%\begin{figure}[t]
  \centering
 \includegraphics[width=0.45\textwidth]{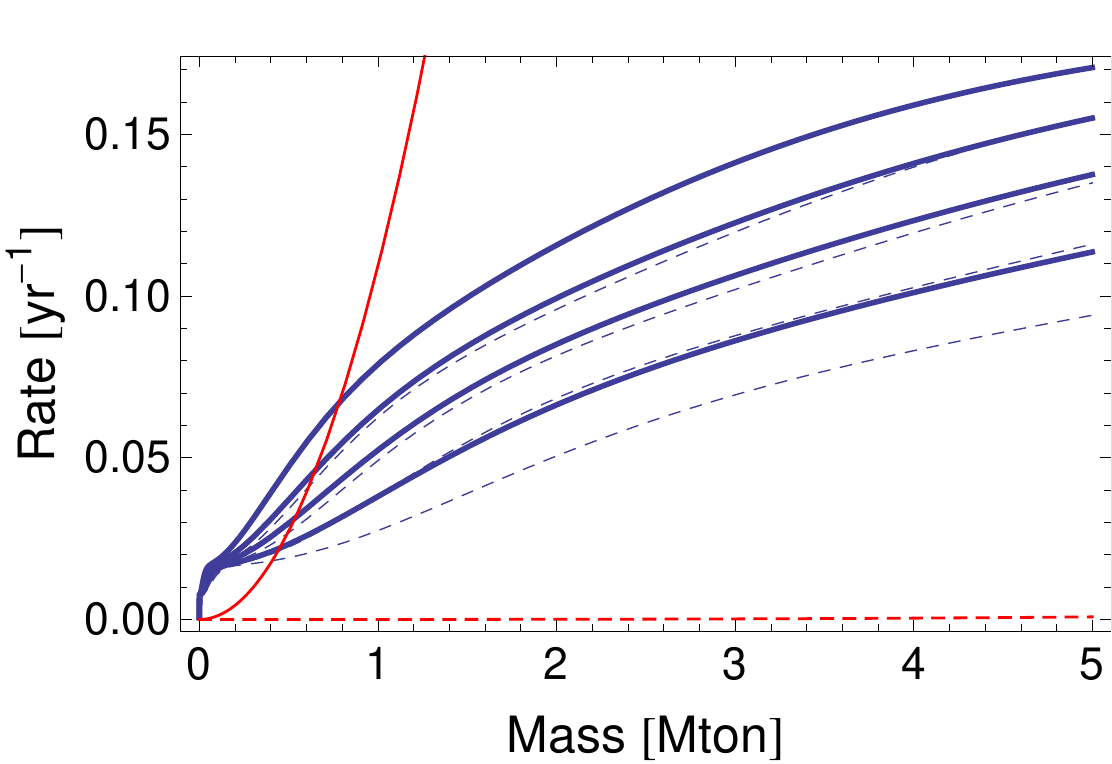}
   \caption{Solid lines: rates of background, $\omega$ (red online), and of detected failed \sn\ bursts, $R^{det}_{BH}$ (blue online), for sources within 10 Mpc distance, as a function of the detector mass, for $N\geq 2$ and  $\bar p=0,0.2,0.4,0.68$ (lower to upper curves).  Note that $\omega \gta R^{det}_{BH}$  for $M\gta 0.8$ Mt.
Dashed  lines: the same results for   $N\geq 3$, for which the background rate is negligibly small (horizontal line).  }
\label{bgr}
\end{figure}
Let us  discuss how our results vary with the parameters.  Rates depend on $f_{BH}$ as $R^{det }_{BH} \propto f_{BH}$ and $R^{det }_{NS} \propto (1- f_{BH})$, so rescaling is immediate.   The dependences on $\bar p$ and on the detector's mass, $M$, are  described in fig. \ref{bgr}, which gives $R^{det }_{BH}$ within a 10 Mpc radius as a function of $M$, for different values of $\bar p$ and for the central curves in fig. \ref{ccrates}b).  For comparison, the background rates are shown; they depend on the mass as $\omega_{2} \propto \lambda^2 \propto M^2$  and $\omega_{3} \propto \lambda^3 \propto M^3$. 
Expectedly, $R^{det }_{BH}$ increases with $M$ and with $\bar p$, due to the increase of the number of events (fig. \ref{spectraevents}) and therefore of the distance of sensitivity. 
Beyond $\sim 1$ Mt of mass $\omega_2 > R^{det }_{BH}$, so  at least three events will probably be needed to establish detection. 
 For a 5 Mt detector like the proposed TITAND \cite{Suzuki:2001rb}, we get $R^{BH}_{det}\simeq 0.10-0.16~{\rm yr^{-1}}$ for $N_{min}=3$.  
 
Results also depend on the equation of state (EoS) of nuclear matter.  For the softer Lattimer and Swesty EoS, the \n\ output of a failed \sn\ is somewhat less luminous and energetic, typically with $E_{0 \bar e} \simeq 20 $ MeV and $L_{0e}\simeq 0.5 \cdot 10^{53}$ ergs \cite{Sumiyoshi:2008zw,Nakazato:2008vj}.  This translates into a reduced distance of sensitivity and therefore a lower rate of detections.  Using the fluxes in \cite{Nakazato:2008vj}, varying over the oscillation parameters and the local \sn\ rate, with M=1 Mt, $f_{BH}=0.22$ and $N_{min}=2$, we find $R^{BH}_{det} \simeq 0.016 - 0.045~{\rm yr^{-1}}$ within 10 Mpc radius. This is close to $R^{NS}_{det} f_{BH}/(1- f_{BH}) $, as expected if the \n\ fluxes were the same in the two collapse types.  
 
Summarizing, the detection of a failed \sn\ at Megaton class \n\ detectors might be a realistic possibility, with a rate of detections reaching about one per decade, depending on the parameters.  This is comparable to the rate of the more common \nts\ collapses, and is due to the larger distance of sensitivity to failed \sne, that includes several major, \sn-rich, galaxies.  The short, $\sim 1$ s duration of a failed \sn\ burst might allow its unambiguous identification already with the coincidence of two inverse beta decay events within this time interval. 

Even with low statistics, the detection of a \n\ burst from a direct \bh\ collapse will have profound implications.   It might be the first observation of a different branch or core collapse, confirming its existence, and giving information on the local rate of failed \sne.   This could be especially interesting in connection with the observed rate of bright \sne\ being lower than expected \cite{Horiuchi:2011zz}, thus allowing for a substantial fraction of failed \sne.
It could also give the exciting  opportunity to witness the formation of a black hole in real time, marked by the sudden truncation of the \n\ burst \cite{Beacom:2000qy}.  Considering the strong dependence of  failed \sn\ \n\ bursts on the equation of state,  conclusions about it might also be possible, with a high rate of \bhf\ bursts favoring a stiffer EoS.

We are grateful to M. Kistler, D. Leonard,  and T. Iida for useful exchanges, and acknowledge the support of the NSF under Grant No. PHY-0854827. 
 
%%%%%%%%%%%%%%%%%%%%%%%%%%%%%%%  

%\bibliography{BHburst}

\end{document}